# The Anatomy of Software Requirements

**Bertrand Meyer*[+], Jean-Michel Bruel[+], Sophie Ebersold[+], Florian Galinier[+], Alexandr Naumchev*[+]**
*Innopolis University
[+]IRIT, University of Toulouse

*Version 3, 11 July 2019. Still a draft. Please check later versions.*

**Abstract**

Requirements engineering is crucial to software development but lacks a precise definition of its fundamental concepts. Even the basic definitions in the literature and in industry standards are often vague and verbose. To remedy this situation and provide a solid basis for discussions of requirements, this work provides precise definitions of the fundamental requirements concepts and two systematic classifications: a taxonomy of *requirement elements* (such as components, goals, constraints…) ; and a taxonomy of possible *relations* between these elements (such as "extends", "excepts", "belongs"…). The discussion evaluates the taxonomies on published requirements documents; readers can test the concepts in two online quizzes. The intended result of this work is to spur new advances in the study and practice of software requirements by clarifying the fundamental concepts.

# 1. Introduction

A software system, like any other engineering construction, exists to satisfy certain human objectives, known as its requirements. The evolution of software engineering has produced ample evidence that the quality of systems fundamentally depends on the quality of their requirements.

It has also led to the realization that *requirements are software*: like code, tests and other products of the software process, requirements for today's ambitious systems are software artifacts, susceptible to some of the same practices (such as configuration management), and in need of theoretical studies. The present discussion defines a standard framework for such studies.

Section 2 explains the scope of the discussion. Section 3 defines basic terminology. The next two sections provide the principal contribution of this work in the form of two taxonomies: a taxonomy of requirement elements themselves in section 4; and a taxonomy of *relations* between requirements in section 5. The rest of the discussion explores the application of these concepts: section 6 applies the taxonomies to analyze an extract from a representative requirements document; section 7 examines popular approaches to requirements engineering in light of the taxonomies; after a discussion of related work in section 8, section 9 assesses the applicability of the approach and prospects for future work, including automatic analysis.

Two online quizzes [Galinier 2019] enable readers to test anonymously their understanding of the taxonomies of requirements and relations.

## 2. Scope

This presentation is *descriptive* rather than prescriptive. Textbooks are an example of prescriptive presentation, stating how one should write requirements. Here the intent is to study requirements as they are, which in the industry's practice does not always mean as they should be. For example, the relationship taxonomy (section 5) has a category for requirements that contradict each other, a case that is obviously not desirable but occurs in practice. Prescriptive discussions will benefit from the analysis, since they should be rooted in a precise understanding of the concepts. Occasionally, as in section 9, the discussion veers into prescriptive territory.

The presentation is, however, *normative*, since it proposes standard definitions and classifications of requirement concepts and terminology relevant to requirements authors regardless of which methodology they follow.

Its ambition is also *universal*: we have tried to cover all possible properties of requirements, with the understanding that this work should be revised if we missed any. In this spirit, enumerations (see for example the list of activities in the definition of "project" in 3.1) never end with such phrases as "etc.", useful to protect authors but detrimental to the quality of definitions. Here there is no such protection; any omission is a mistake and will have to be corrected.

While section 9 is the place for a more detailed analysis of the applicability of this work, it is legitimate to ask at the outset for a general justification: why is it worthwhile to engage in such an effort at precision (at the risk of pedantry) to define and classify concepts that are widely used in practice with their intuitive meaning?

The general justification is that requirements are a difficult concept to apprehend because they straddle the border between the formal and the informal, the exact and the approximate, the technical and the human. Some software engineering concepts are formal, exact and technical: programming languages, for example, have precise definitions, and any single detail of a million-line program may critically affect its correctness. At the other end of the spectrum, equally important concepts of software engineering, such as methods of project management, are informal, approximate and human.

Requirements bridge these two worlds. To be effective, they must cover the needs of both. Insufficient rigor in the handling of requirements concepts hampers this goal. As an example, there is wide disagreement in the field as to what constitutes the difference between "functional" and "non-functional" requirements, to the point that some authors even reject this distinction altogether. The rest of the literature treats it as a given, but without a generally agreed precise definition.

A software system is often just one part of a larger system whose other elements may be people and organizations, as in enterprise systems, or physical devices, as in cyber-physical systems. While the authors' primary interest and the examples in this article are software-related, the intent of the definitions and taxonomies is to encompass systems of any kind.

## 3. Underlying concepts

To discuss requirements we need a set of basic concepts and their precise definition. This section introduces the terminology that serves as a basis for the rest of the discussion. It does not intentionally introduce any novel concept, but gives precise definitions of known concepts. These definitions are not the most general possible ones for the corresponding English words as used in an arbitrary context; rather, they are tailored to the needs of this discussion of requirements.

## 3.1 General concepts

**Universe of discourse**. The assumed context for the present discussion is a ***project*** to develop a ***system*** in a certain ***environment***.

*Comment*: the definitions of project, system and environment follow.

**Definition**. A ***system*** is a set of related artifacts.

*Comment*: In the case of pure *software* systems, the artifacts are virtual: programs, databases, design diagrams, test cases… In line with the goals stated in section 2, the definition is more general, encompassing enterprise and cyber-physical systems. Even if the system involves only software, the project and the environment may include material and human elements.

**Definition**. A ***project*** is the set of human processes involved in the planning, construction, revision and operation of a system .

*Comments*:

- A project is, per this definition, applied to one system. While a project can in practice involve the development of several systems, the definition loses no generality since we can consider them, for the purpose of the definition, to be subsystems of one larger system.
- A particular project may involve only some of the activities mentioned (planning, construction, revision, operation). In particular, the revision of a system (which may also be called maintenance, reconstruction, redesign, evolution and "brownfield development") can be an extension of a previous project for this system, or a new project.

**Definition**. An ***environment*** for a project or system is the set of entities (people, organizations, regulations, devices and other material objects, other systems) external to the project or system but with the potential to affect it or be affected by it.

*Comment*: the environment is also called, in classic Jackson-Zave terminology [Jackson 1995], the "domain". It includes all external elements constraining the project or the system; "external" in the sense that unlike features of the system and project they are imposed from the outside and not susceptible to decisions by the project. As an example of the difference, *"all accounts must maintain a non-negative balance at all times"* is an environment property (affecting the system); "*a withdrawal request for an amount greater than the balance shall produce an error message and leave the balance unchanged*" is a system property, devised to enforce the preceding environment property. Similarly, "*at least 50% of the code shall be developed in-house*" is an environment property (affecting the project); "*the implementation of the user interface module shall be outsourced to company X*" is a project property, which should comply with the environment property.

Environment properties in requirements will be called *constraints* (4.1.B).

## 3.2 Properties and their statements

The definition of "requirement" will use the auxiliary concept of "statement", itself relying on the notion of "property" (a term already used informally). These are general term, not specific to software or requirements; although they essentially retain their ordinary meaning, it is useful for the purposes of the present work to give them precise, slightly more restricted definitions.

**Definition**. A ***property*** is a boolean predicate.

*Comment*: an example of property is that today is Sunday, a predicate (true or false in a given context). The properties of interest for this discussion will apply to a project, system or environment. A system example is the property that response time for a certain kind of query must not exceed one second. A project example is the property that the project uses sprints (iterations) of one month each. An environment example is the property that no more than 50 vehicles at a time are permitted in a tunnel.

**Definition**. A ***statement*** is a human-readable expression of a property.

*Comments*:

- Discussions of programming languages use the term "statement" to mean "instruction", a command to be executed by a computer (prescriptive). Instead "statement" as used here retains the same connotation as in ordinary English: a phrasing that "states" a property (descriptive).
- "*Today is Sunday*" and "*query response time shall not exceed one second*" are statements. The difference between a property and a statement is that the property is the abstract predicate and a statement its expression in a certain notation. Different statements can express the same property; for example the statement "*c'est aujourd'hui dimanche*" is a different statement (in French) of the first example's property.
- A statement, however, specifies just one property. This convention causes no loss of generality since a property, being a predicate, can be built out of logical combinators such as "and" and "or", and hence arbitrarily complex. The next definition will reflect this observation.
- Not all statements have to be expressed, like the preceding examples, in natural language: a statement could be a UML diagram specifying a system property, a mathematical formula describing a constraint property, a PERT diagram or (in agile development) a burndown chart specifying a task property. For any statement, it should be clear what underlying notation it uses (see the notion of "requirement type" in 3.5).

**Definition**. A property, and a statement expressing it, are ***composite*** if the property is a logical combination of simpler properties, and ***elementary*** otherwise.

*Comment*: since a property is a boolean predicate, it may result from applying boolean operators to one or more simpler properties, in which case we call it composite.

**Definition**. A composite property, and a composite statement expressing it, are ***homogeneous*** if the property combines properties of a similar nature, and ***heterogeneous*** otherwise.

*Examples*: "customers will have access to customer functions, and employees to both customer and flight management functions" (from [Bandakkanavar 2017]) is homogeneous. "Error messages shall be recorded in a log" specifies both the presence of a system component (if the log is not defined elsewhere) and a system behavior, and hence heterogeneous.

*Comment*: from a prescriptive viewpoint (as discussed in section 2), it is good practice for requirements documents to avoid heterogeneous statements. The second example would be better expressed, in a requirements document, by two distinct requirements: one specifying the need for a log; the other stating that error messages must be recorded in that log.

### 3.3 Relevant properties

The definitions of "property" and "statement", when applied to projects and the associated system and environment, underlie the definition of "requirement". But many properties are not of interest as requirements, for example the system property that the executable has a "load" instruction at offset 3FD04, or the project property that no code was committed past 11:30 PM on December 31st. We are interested in properties that are relevant to some stakeholder.

**Definition**. A ***stakeholder*** for a project is a person who may affect or be affected by the project or its associated system.

*Comments*:

- This definition is a considerably simplified version of the one on the IEEE systems and software terminology standard [IEEE 2010]. The IEEE version talks of a person or *organization*, but organizations can only be involved through their (human) members. It specifies "*individual or organization having a right, share, claim, or interest in a system or in its possession of characteristics that meet their needs and expectations*", "*individual, group or organization that can affect, be affected by, or perceive itself to be affected by, a risk*" etc., all possibly interesting but only adding musings to the simple definition above. (The mention of *perceiving* to be affected is correct but not necessary: if you believe you are affected by the system you are affected by it, if only through the effect on your mindset.) There seems to be no need for such a bloated definition for a clear and simple concept.
- Concretely, stakeholders may include users of the system, people responsible for commissioning and accepting the system (such as "product owners" in agile methods), developers, testers and many others as discussed in detail in the software engineering literature, e.g. [Laplante 2013].
- The definition only mentions the *project* and *system*. Affecting or being affected by the *environment* is not enough to make you a stakeholder. As a taxpayer you are affected by the tax rules, but that does not make you a stakeholder of a tax-related project if the resulting system does not apply to your category of taxpayer.

**Definition**. A property of a project, system or environment, is ***relevant*** if it is of interest to a stakeholder.

*Comment*: we saw above examples of non-relevant project and system properties. As an example of an environment property, knowing that the system might be deployed in Costa Rica is relevant for a payroll system which must take local regulations into account, but probably not for a computer game.

### 3.4 Requirement

**Definition**. A ***requirement*** is a statement of a relevant project, system or environment property.

*Comments*:

- This definition introduces the central concept of the present discussion. From the definition of "statement", a requirement is a specification of a property of a project, system or environment. (For simplicity we limit ourselves to requirements characterizing only one of the three dimensions.)
- The classification of section 4 defines what kinds of property are pertinent for software requirements.
- Software engineering discussions often use the plural "requirements" as a collective, as in "the requirements of a system", a phrase that denotes a whole (the specification of the system) beyond just the collection of its parts (the individual requirements). To avoid any ambiguity, the present discussion only uses "requirements" as the plural of "requirement", as in "four requirements", meaning four statements of project, system or environment properties. For the collective we can always use a more elaborate phrase such as (depending on the exact meaning sought) "the requirements document" or "the overall requirements for the system", or "the Software Requirement Specification", often abbreviated SRS.
- The definition only says that a requirement specifies a property, and does not specify a level of granularity for that property: it could characterize the entire project, system or environment, one of its major components, or just an elementary component. At one extreme, the entire SRS is "a requirement"; so is, at the other extreme, the statement of a single elementary property, such as "Clicking Exit shall result in termination of the session". The next definition addresses this variety.
- By specifying a boolean property, a requirement defines a criterion which an actual environment, project or system either confirms or refutes. "*Have the test plan ready for next Monday!*" is not boolean and hence not a requirement. ("*The testing team shall produce the test plan in at most a week*" is a requirement.) When teaching requirements engineering we go further, telling students that requirements must be verifiable: "*the query shall be processed in real time*" is not good enough, "*query response time shall be one millisecond or less*" is better (see e.g. [Wiegers 2014] for such advice). Here again, the present document is descriptive and taxonomic, not normative. Except in section 9 it does not discuss what makes requirements "good", only what makes them requirements.

### 3.5 Characterizing requirements

**Definition**: A requirement is ***composite*** if it includes other requirements (its ***sub-requirements***) and ***elementary*** otherwise.

*Comment*: the distinction is the same as for "statements" in general (3.2) but introduces the notion of sub-requirement, which will become more precise through the definitions of "component", "sub-goal" etc. in section 4.

**Definition**. The ***type*** of a requirement is the notation in which it states its associated property.

*Comment*: the term "notation" is taken here in its ordinary meaning. Examples of notation are English text, a UML diagram type, a tabular format, a particular programming language, a (well-defined) mathematical notation. Since requirements can be composite, the notion of "notation" must support the possibility of a combination of notations, as in the example of a requirements document that contains both English text and graphical illustrations.

**Definition**. A requirement R ***specifies*** a property P if P follows from the property stated by R or a sub-requirement of R.

*Comment*: this definition is a bit of hair-splitting but reflects the different nature of statements and properties. A property is just a predicate: the border of a certain control on the screen is (or is not) black. A statement is an expression of that property in some notation, for example "*The border shall be black*" or "*La bordure doit être noire*", both of which express the same property although in different notations (types). Yet another way to specify that property would be a figure, or an entry in a table listing attributes of UI elements. The definition uses the informal term "follows from" since it cannot use "R implies P" unless requirements are expressed in a formal mathematical notation.

# 4. Classification of requirements

This section introduces the first of the two fundamental taxonomies proposed by this article: the taxonomy of requirements themselves. Section 4.1 defines the fundamental categories, disjoint from each other. Section 4.2 introduces other categories, important in practice but defined as subcategories of the fundamental ones.

**4.1 Requirements classification: basic categories**

**Classification**. Every requirement states a property of one of the following categories. Section 4.2 will introduce more categories as special cases of the fundamental ones given here.

A. **Component:** the property that the system, project or environment includes a certain part.
   *Comment*: a component can be material, virtual or human. A human component can be a single person, group of persons, organization or category of persons involved in the system, project or environment. A component of the environment can be another system with which the given system must be interfaced.
   *Examples*: "the operating system is designed to run on the iPhone 8 and later models" (system component, material); "database operations shall run in a separate process" (system component, virtual); company CEO (if referenced explicitly in the requirements, single person); reservation agents (category of persons).

B. **Goal**: an objective of the project or system, in terms of their desired effect on the environment.
   *Example*: "One of the advantages expected from the system is to reduce the amount of fraudulent invoices".
   *Comments*: Requirements documents often present goals at the beginning of the text. The external entity could be a company (enterprise goals) or a physical device such as a phone (cyber-physical goals). Having an effect on the environment means having an effect on an external entity, such as a company (enterprise goals, as in this example) or a physical device (cyber-physical goals).

C. **Behavior**: a property of the results or effects of the operation of the system or some of its components.
   *Example*: "Display the list of available elements."
   *Comments*: requirements in this category often get the most attention since they describe elements of what the system will do. A behavior can characterize the system as a whole or a specific component. Section 4.2 introduces the classic distinctions of behaviors into functional and non-functional.

D. **Task:** the property that the project includes a certain activity.
   *Examples*: program coding, stakeholder interview, daily meeting.

E.   **Product**: the property that a task uses or produces a material or virtual object.
     *Examples*: a test plan, a user story, a design document, a program module.
F.   **Constraint**: an environment property that may affect components, goals, behaviors, tasks or products.
     *Examples*: "every transfer over EUR 10,000 requires authorization" (behavior constraint); "testing shall use the JUnit framework" (task constraint).
     *Comment*: it would seem enough to say "an environment property", since by definition the environment is (3.1) the set of external entities that have the potential to affect or be affected by the project (and hence the system and the environment). But this does not work, since those entities have other properties with no relation to the project. Hence the restrictive formulation. 4.2 will distinguish between *obligation* and *assumption* constraints.
G.   **Role**: the property that a component carries some or all of the responsibility for a behavior or task.
     *Examples*: "the Bangalore subsidiary shall be responsible for the implementation of the user interface subsystem" (task role, human component of the project); "the reservation system's UI shall be designed for operation by railway-station booking agents" (behavior role, human component); "smart contract computations shall be executed on the GPU" (behavior role, material component).
H.   **Limit**: the property that the project, system or environment does *not* include a requirement of one of the preceding kinds.
     *Examples*: "Providing a interface to SAP accounting falls outside of the scope of the present system" (component limit);  "Integration testing will be performed in a follow-up project (project limit).

I.   **Lack**: a property that should have a requirement, but does not.
     *Comment*:  this category is different from the others, and paradoxical since it characterizes what is *not* in the requirements. Our discussions with requirements practitioners indicate that they spend a considerable part of their efforts uncovering  lacks. Human scrutiny is indeed usually required to find lacks, although some automatic analysis is possible; for example, a term that appears repeatedly in an SRS but not as an entry in the glossary (a list of definitions of project, system and environment concepts, which any SRS should include) may signal that the requirements are missing the specification of an important property.

J.   **Meta-requirement**: a property of requirements themselves (not the system, project or environment).
     *Examples*: a section title in the requirements document (which does not express any new property but helps structure and understand the actual, non-meta properties); more generally, any observation intended to facilitate the reading of an SRS, such as  "the details will appear in section 7"; a statement of priority between requirements, such as a classification of components into "critical", "necessary" and "nice to have"; an explanation, such as "the behavior in this case is specified by table 7.1" or "figure 7.2 illustrates the concept".

*Comment*: large composite requirements, for example an entire SRS, will contain requirements in several of these categories. The classification is, however, designed with the intent that in practical usage it will be possible without much hesitation to classify any elementary requirement (or small composite requirement) into just one category.

The classification makes it possible to be more precise about the elements of a composite requirement (a requirement made of other requirements):

**Definition**: a *sub-goal*, *sub-component*, *sub-behavior* etc. is a sub-requirement of respectively a goal, component, behavior etc.

And consequently:

**Definition**: A goal, component, behavior etc. is *elementary* (non-composite) if it has no sub-goal, sub-component, sub-behavior etc.

*Comment*: in principle, the definition of sub-requirement allows arbitrary mixing of categories, for example a task as a sub-requirement of a goal. The above definitions only cover sub-requirements that are of the same category as the enclosing requirements.

### 4.2 Some derived categories

The following kinds of requirement are special cases, important in practice, of the categories of section 4.1.

An **actor** is a human *component*. Examples include the stakeholders of a project as defined in section 3.3 (project actors); and people involved in the operation of the system, such as an end-user or a system administrator (system actors).

A **justification** is a *meta-requirement* explaining the rationale for a requirement (of any kind) in terms of a *goal*. As an example, if an SRS for a software system does not specify Android among the platforms to be supported, it might include the justification that the company has made the strategic decision to equip its sales agents with iPhones.

A **responsibility** is a human *role*. (In the general case, roles can be defined for components other than humans, e.g. software components.) The first two examples in the above definition of "role" (4.1, H) are responsibilities.

An **obstacle** is a *goal* defined as the need to overcome a negative property of the environment, as in "with the current system, too many enquiries that could lead to sales are missed". KAOS [van Lamsweerde 2000] has a closely related definition.

A widely established terminology for *behavior* distinguishes between statements of "what" and "how" properties:

- A **functional behavior** specifies results or effects of the system.
- A **non-functional behavior** specifies a property of how these results or effects are to be achieved. Classical examples are timing limits and security conditions.

The following subcategories exist for *constraints* (environment properties):

- A **business rule** is a constraint resulting from organizational practices. *Examples* are the rules on bank accounts from 3.1 and 4.1.F. Another is "delivery of phosgene [a chemical] requires that the recipient have taken a refresher course in handling hazardous chemicals in the past twelve months". This example appears in [Wiegers 2014], as the background for a system property: the software must reject a request for chemical if the requester does not meet the criterion.

- A **physical rule** is a constraint resulting from laws of nature. A typical *example* is the application of the laws of mechanics to a satellite launching system.
- An **engineering decision** is a constraint resulting from human choices. *Examples* are the minimum and maximum bandwidths for a networking system.

A separate classification of *constraints* is between:

- An **obligation**, describing a property that the environment is known to possess. *Examples*: the rules on bank transfer in 4.1.E; in a cyber-physical system, limits (such as signal transmission speed, laws of mechanics, bandwidth) imposed by physics and engineering.
- An **assumption**, describing a property that the environment may or may not possess but which the project may assume to hold for the development of the system. *Example* (in a system to control a railroad crossing): "cars travel at no more than 200 km/h and trains at no more than 400 km/h". Unlike the absolute limits imposed by the laws of nature or by a choice of technology, an assumption is the result of an explicit human decision, and might conceivably not hold, but may be needed for the requirements to guarantee certain properties. In the example, it may be possible to make trains run faster than 400 km/h, but no railroad-crossing system can guarantee the avoidance of collisions without assuming some upper limit on the speed of trains.
- An **invariant**, describing a property that is both as an assumption and as a behavior. *Example* (in a factory control system): "the system shall operate between -30 and +50 degrees Celsius", which means both that the system's operations *may* assume they start within this temperature range and that they *must* refrain from causing overheating or over-cooling.

While requirements of all three kinds cover properties of the environment, the difference is important in practice since obligations make the work of system developers harder and assumptions make it easier. (Invariants do both. To keep the three categories disjoint we classify a constraint as an obligation if it is not also an assumption, and conversely.)

The two classifications are orthogonal: for example, a business rule can be an obligation (as the bank transfer example rule) or an assumption (the New York Stock Exchange is closed on Labor Day). The same observation holds for engineering decisions, which gave us an example of obligation (car and train speeds) and an example of invariant (temperature limits).

The following table, intended for reference, includes all the categories in alphabetical order, and their subcategories. Every requirement should fit into exactly one category and at most one subcategory (except for constraints which may belong to elements of the two orthogonal classifications).

| Basic categories | Subcategories | Short definition (for full definition see text) |
|---|---|---|
| **Behavior** | | Property of an operation's effects |
| **Component** | | Part of the project, environment or system |
| | *Special case*: **Actor** | Human component |

| Constraint | | | Environment property |
|---|---|---|---|
| | *Classification by nature*: | **Assumption** | Assumed constraint |
| | | **Obligation** | Imposed constraint |
| | | **Invariant** | Both assumption and obligation |
| | *Classification by source*: | **Business rule** | Constraint due to organizational practices |
| | | **Engineering decision** | Constraint due to human choices |
| | | **Physical rule** | Constraint due to laws of nature |
| **Goal** | | | Intended effect of project or system on environment |
| **Lack** | | | Missing requirement |
| **Limit** | | | Property beyond scope of project/system/environment |
| **Meta-requirement** | | | Property of requirements but not of project, system or environment |
| | *Special case*: **Justification** | | Rationale expressed in terms of a *goal* |
| **Product** | | | Material or virtual object used or produced by a task. |
| **Role** | | | Component's responsibility for behavior or task |
| **Task** | | | Project activity |

An anonymous online quiz [Galinier 2019-1] invites readers to test the practicality of the requirements classification and their understanding of it by classifying requirements elements from a sample requirements document [Bair 2006], which also provides the background for the discussion in section 6.

## 5. Taxonomy of inter-requirements relations

With requirement elements sorted into categories, we proceed to a classification of the relations that may hold between them.

**Classification**. A requirement Y may depend on another X in one of the following ways, each given with: a name in upper case (a verb, such as "REPEATS", whereas names of requirement categories were nouns); a symbol (generally borrowed from mathematics, for its mnemonic value only); a definition of its meaning; a comment if necessary.

DISJOINS          $\mathtt{X \;||\; Y}$      Y and X are unrelated.
*Comment*: In this case, the most common for two randomly selected statements in a requirements document, there is no relation between the properties they specify.

BELONGS          $\mathtt{X \subseteq Y}$      X is a sub-requirement of Y.
*Comment*: this case corresponds to textual inclusion (sub-section, sub-figure etc.), unlike inclusion of *properties* as in EXTENDS below.

REPEATS          $\mathtt{X \Leftrightarrow Y}$      X specifies the same property as Y.
*Comment*: this case is identity of the properties although not necessarily of their statements (since they might use different notations). See below for variants: EXPLAINS (different notations), DUPLICATES (same notation).

CONTRADICTS          $\mathtt{X \oplus Y}$      X specifies a property in a way not compatible with Y.
*Comment*: remember that this discussion is descriptive, not prescriptive. No one would *recommend* writing contradictory requirements. But existing SRS, especially large ones, may contain contradictions; in some contexts it might be crucial to detect them.

FOLLOWS          $\mathtt{X \dashv Y}$      The property specified by X is a consequence of the property specified by Y.
*Comment*: interesting in particular if Y is a goal and X a behavior or task.

EXTENDS          $\mathtt{X \;>\; Y}$      X assumes Y and specifies a property not specified by Y.
*Comment*: also called "refines".

EXCEPTS          $\mathtt{X \;\backslash\backslash\; Y}$      X changes or removes, for a specified case, a property specified by Y.
*Comment*: this case is not the same as CONTRADICTS. It is the explicit and often legitimate introduction of an exception to a general property.

CONSTRAINS          $\mathtt{X \blacktriangleright Y}$      X specifies a constraint on a property specified by Y.

CHARACTERIZES          $\mathtt{X \rightarrow Y}$      X is a meta-requirement involving Y.

The following derived cases are useful in practice:

DETAILS          $\mathtt{X \;\gg\; Y}$      X adds detail to a property specified by Y.

|            |       |                                                                                          |
|------------|-------|------------------------------------------------------------------------------------------|
|            |       | *Comment*: this is a case of X > Y (EXTENDS). The nuance is that in this case there is no fundamentally new property, just more detail about an already specified property. |
| SHARES     | X ∩ Y | X' ⇔ Y' for some sub-requirements X' and Y' of X and Y. (Involve REPEATS.) |
| DUPLICATES | X ≡ Y | X ⇔ Y, and X has the same type as Y. *Comment*: also a case of REPEATS. This is the true redundancy case. From a prescriptive viewpoint, it usually reflects a deficiency in an SRS; compare with the next case. |
| EXPLAINS   | X ≅ Y | X ⇔ Y, and X has a different type from Y. *Comment*: again a case of REPEATS, but not necessarily bad. Y introduces no new property but helps understand Y. For example Y may describe a property textually, and X may be a graphical illustration of that property. |

*Comments*:

- As with the taxonomy of requirements, the intent is to ensure that given two arbitrary requirement elements their relationship can be classified in at most one of the primary relations and at most one of the derived ones. If two or more categories appear to apply, one should clearly be more relevant than the others.
- The mathematical symbols informally suggest the relations' meaning, but do not imply the properties, such as associativity or commutativity, of their ordinary mathematical counterparts. Further research should indeed study (in the style of [Meyer 1985-1]) the mathematical properties of these relations.
- The relations may hold between requirements of any complexity. In practice, one should first look for their occurrences between elementary requirements.
- SHARES is an example of a relation on composite requirements derived from another (DUPLICATES) on their sub-requirements. It is possible to generalize some of the other relations in the same way, or simply to accept, as a small abuse of language, that for example Y > X holds if Y' > X' holds for sub-requirements. Except for SHARES, we ignore this issue in light of the preceding comment.
- An analysis examining how two given requirements are connected may in principle identify more than one of the relations. For simplicity, it is advisable to choose only one (from the complete list including derived relations); just pick the relation that comes out as most relevant.

Like its counterpart for the first taxonomy, the following table provides a list of all the categories and subcategories of the relation taxonomy.

| *Basic categories* | *Subcategories* | *Symbol* | *Short definition (for full definition see text) Y is first operand, Y second operand* |
|---|---|---|---|

| BELONGS | | ⊆ | X textually included in Y |
|---|---|---|---|
| CHARACTERIZES | | → | Meta-requirement X applies to Y |
| CONSTRAINS | | ▶ | Constraint X applies to Y |
| CONTRADICTS | | ⊕ | Properties specified by X and Y cannot both hold |
| DISJOINS | | \|\| | X and Y are unrelated |
| EXCEPTS | | \\ | X specifies an exception to the property specified by Y |
| EXTENDS | | > | X adds to properties of Y |
| | *Special case*: **DETAILS** | » | X adds detail to properties of Y |
| FOLLOWS | | ⊣ | X is a consequence of Y |
| REPEATS | | ⇔ | X specifies the same property as Y |
| | SHARES | ∩ | Some subrequirement is common |
| | DUPLICATES | ≡ | Same properties, same type (notation) |
| | EXPLAINS | ≅ | Same property, different type |

As with the previous taxonomy, an anonymous online quiz [Galinier 2019-2] invites readers to test the practicality of the requirements-relations classification and their understanding of it by classifying requirements relations from a sample requirements document [Bair 2006], which also provides the background for the discussion in section 6.

# 6. Dissecting an example

[Blair 2006] is an example requirements document, obviously inspired by industrial practice but devised for a course at Ohio State University. It provides a good testbed for the concepts of this article since it is small enough to lend itself to analysis yet large and realistic enough to be representative of the contents of requirements for actual industry projects.

We analyzed the entire text and found that the taxonomies cover both all requirements and all the relations we considered. Here we only show a few representative samples of the analysis. The entire analysis is available as an online complement to this article [Galinier-2019-3].

First, examples of classifying requirements according to the first taxonomy:

| Section 1. Introduction | Meta-requirement |
|---|---|
| 1.1 Purpose of Document | Meta-requirement |
| This is a Requirements Specification document for a new web-based sales system for Solar Based Energy, Inc. (SBE) | Goal |
| 1.2 Project Summary | Meta-requirement |
| Project Name:   SBE Sales System | Component |

1.4 Project Scope

| The scope of this project is a web-based system that supports the marketing of SBE products directly to customers as well as through the existing sales agent network. | Goal |
|---|---|
| Advertising of products, inventory control, and account billing are not part of this project. | Limit |
| In addition, changes to the logical and physical design of the current databases are expected. | Obstacle |
| The primary responsibilities of the new system: | Meta-requirement |
| provide customers direct access to up-to-date, accurate product information on which they can make a decision to buy | Behavior |

Section 2. Functional Objectives

| 2.1. High Priority | Meta-requirement |
|---|---|
| "The system shall allow for on-line product ordering by either the customer or the sales agent" | Behavior |
| "For customers, this will eliminate the current delay between their decision to buy and the placement of the order" | Goal |
| "This will reduce the time a sales agent spends on an order by x%. The cost to process an order will be reduced to $y" | Goal |
| "The system shall display information that is customized based on the user's company, job function, application and locale" | Behavior |
| 2.2 Medium Priority | Meta-requirement |
| The system shall provide a search facility that will allow full-text searching of all web pages that the user is permitted to access. | Goal |
| The system must support the following searches:<br>    * find all words specified<br>    * find any word specified<br>    * find the exact phrase<br>    * Boolean search | Behavior |

Section 3: Non-Functional Objectives

| | |
|---|---|
| 3.1" Reliability" | Meta-requirement |
| * "The system shall be completely operational at least x% of the time" | Constraint |
| * "Down time after a failure shall not exceed x hours" | Constraint |

Section 4: The Context Model

| | |
|---|---|
| 4.1 "Goal Statement" | Meta-requirement |
| "The goal of the system is to allow SBE to increase sales revenue by x% over the next y years with only a z% increase in sales and customer service staff by" | Goal |
| * "allowing complete and accurate customer and order information to be captured directly from the customer as well as from sales agents" | Goal |
| 4.2 "Context Diagram" | Meta-requirement |
| [Context diagram showing SBE Sales System with external entities: Customer, Sales Agent, Product Owner, Accounting, Shipping, Marketing, with data flows including Keyword Matches, Product Information, Whitepapers, Customer Data, Preferences, Orders, Customer Request, Customer Navigation Paths, Purchase Transaction, Order Details] | Behavior |

| 4.3 "System Externals" | Meta-requirement |
|---|---|
| "Customer" | Actor |
| "A customer is any user of the system that has not identified himself as an SBE employee" | Actor |
| "A customer may search for public product information by keyword, access whitepapers for a particular product, order a product or request assistance from a sales agent" | Role |
| "A customer who provides personal information will get search and query results customized to his preferences" | Behavior |

5. The Use Case Model
 5.1 System Use Case Diagram

| 5.2 Use Case Descriptions (for selected cases) | Meta-requirement |
|---|---|
| * "For all use cases, the user can cancel the use case at any step that requires user input. This action ends the use case. Any data collected during that use case is lost" | Behavior |
| * "For all use cases that require a logged in user, the current login session is updated during the use case to reflect the navigation paths through the use case" | Behavior |
| Use Case Name: Login User | Meta-requirement |
| Summary: In order to get personalized or restricted information, place orders or do other specialized transactions a user must login so that the system can determine his access level | Goal |
| Basic Flow | Meta-requirement |
| 1. The use case starts when a user indicates that he wants to login. | Constraint |
| 2. The system requests the username and password. | Behavior |
| 3. The user enters his username and password. | Role |
| 4. The system verifies the username and password against all registered users. | Behavior |
| | |
| Alternative Flows | Meta-requirement |
| Step 4: | |
| if username is invalid, the use case goes back to step 2. | Behavior |
| | |
| Extension Points: none | Component |
| Preconditions: The user is registered. | Constraint |
| Postconditions: The user can now obtain data and perform functions according to his registered access level. | Behavior |

| Business Rules: Some data and functions are restricted to certain types of users or users with a particular access level" | Constraint |
|---|---|

Now, some examples of requirements relationships per the second taxonomy.

CONSTRAINS:

| "Preconditions: The user is registered." | "Postconditions: The user can now obtain data and perform functions according to his registered access level." |
|---|---|

EXCEPTS:

| "if the password is invalid the system requests that the user re-enter the password. When the user enters another password the use case continues with step 4 using the original username and new password." | *"4. The system verifies the username and password against all registered users".* |
|---|---|

BELONGS:

| "A customer is any user of the system that has not identified himself as an SBE employee." | "4.3 System Externals<br>Customer<br>A customer is any user of the system that has not identified himself as an SBE employee. A customer may search for public product information by keyword, access whitepapers for a particular product, order a product or request assistance from a sales agent. A customer who provides personal information will get search and query results customized to his preferences.<br>Sales Agent<br>A sales agent is a user who has been verified as an SBE employee. A sales agent may access all available product information and whitepapers, including the product owner. A sales agent may place an order on behalf of a customer. He will be informed by the system of any customers in his region who have requested assistance.<br>Product Owner<br>The product owner is a user who has been verified as an SBE employee. The product owner may update product information and whitepapers for those products for which he is responsible.<br>Accounting<br>The Accounting department is responsible for all SBE financial transactions. The Accounting department is informed of all purchases and is responsible for later collection of accounts |
|---|---|

|  | receivable.<br>Shipping<br>The Shipping department is informed of purchases so that it can process the order and update inventory.<br>Marketing<br>The Marketing department is responsible for creating demand for SBE products. It will receive website navigation data to use in planning marketing strategies." |
|---|---|

DETAILS:

| "The system shall be completely operational at least x% of the time" | "Down time after a failure shall not exceed x hours" |
|---|---|

CHARACTERIZES:

| "2.1 High Priority" | "The system shall allow for on-line product ordering by either the customer or the sales agent." |
|---|---|

DISJOINS:

| "A sales agent may access all available product information and whitepapers, including the product owner. A sales agent may place an order on behalf of a customer" | "if the password is invalid the system requests that the user re-enter the password. When the user enters another password the use case continues with step 4 using the original username and new password." |
|---|---|

EXPLAINS:

| [Data flow diagram showing SBE Sales System with Customer, Sales Agent, Product Owner, Accounting, Shipping, and Marketing entities] | "The goal of the system is to allow SBE to increase sales revenue by x% over the next y years with only a z% increase in sales and customer service staff by<br>- allowing complete and accurate customer and order information to be captured directly from the customer as well as from sales agents<br>- providing customers and sales agents fast access to up-to-date and accurate product information and whitepapers." |
|---|---|

# 7. Analyzing available requirements methodologies (draft section)

This section surveys a few important requirements methodologies, selected from those covered in a recent survey involving some of the authors [Bruel 2019].

 At this stage we only consider the classification of requirements in well-known requirements textbooks.

## 7.1 Wiegers-Beatty

Wiegers and Beatty ("WB"), include in [Wiegers 2014], page 7, a table of requirements categories, with the following figure (page 8) illustrating their connections:

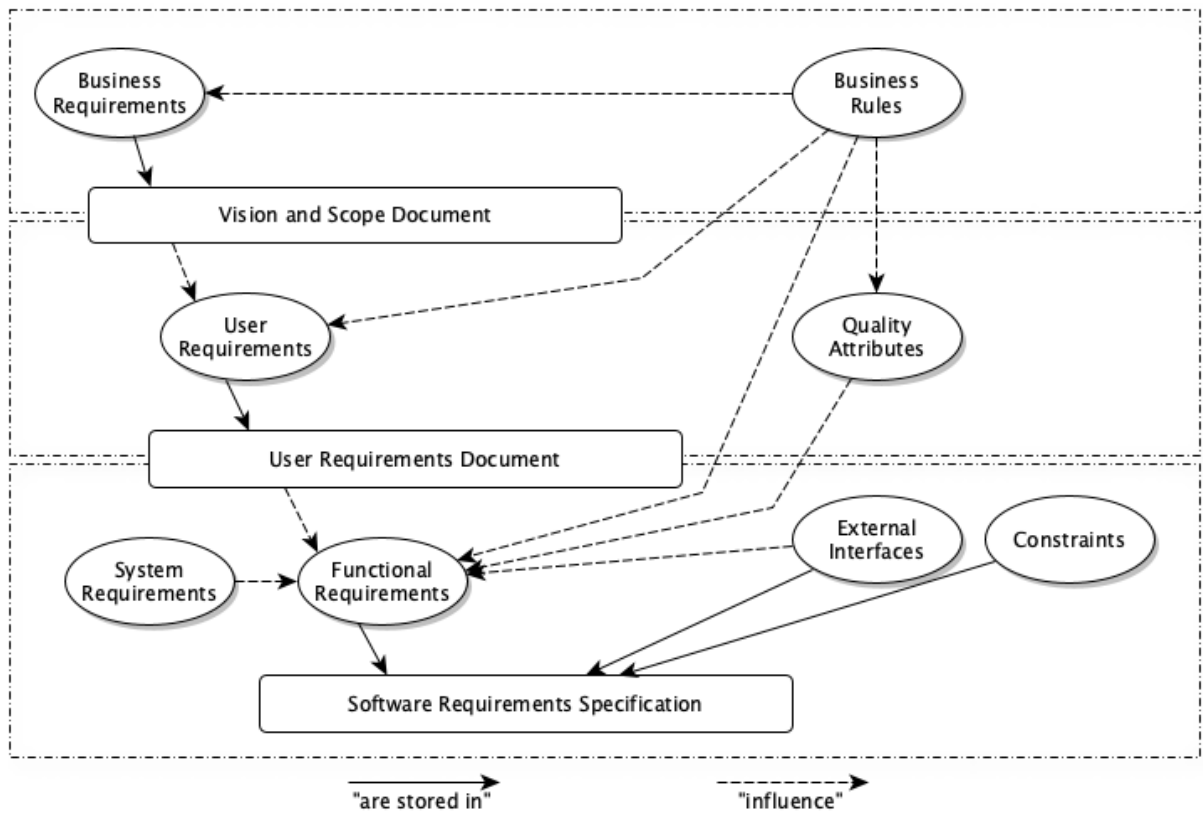

Categories and relationships (from Wiegers 2014)

The first two columns in the following table are reproduced from Wiegers and Beatty; the third column gives in each case the corresponding category in the present classification.

| WB category | WB definition | Category from the present discussion | Comment |
|---|---|---|---|
| Business requirement | A high-level business objective of the organization that builds a product or of a customer who procures it. | *Goal* | Can also include *limits* |
| Business rule | A policy, guideline, standard, or regulation that defines or constrains some aspect of the business. Not a software requirement in itself, but the origin of several types of software requirements. | *Constraint* | See also *business rule* subcategory (4.2) |
| Constraint | A restriction that is imposed on the choices available to the developer for the design and construction of a product. | *Constraint* on *behavior* or *Task* | |
| External interface requirement | A description of a connection between a software system and a user, another software system, or a hardware device. | *Component* | |
| Feature | One or more logically related system capabilities that provide value to a user and are described by a set of functional requirements. | *Behavior* | From viewpoint of a*ctor* (e.g. user) |
| Functional requirement | A description of a behavior that a system will exhibit under specific conditions. | *Behavior* | |
| Nonfunctional requirement | A description of a property or characteristic that a system must exhibit or a constraint that it must respect. | *Constraint* on the *system* or *products* | |
| Quality attribute | A kind of nonfunctional requirement that describes a service or performance characteristic of a product. | *System constraint* (Note: not clear what the difference is with the previous category) | From viewpoint of a*ctor* (e.g. user) |
| System | A top-level requirement for a | *Component* | |
| requirement | product that contains multiple subsystems, which could be all software or software and hardware. | | |
| User requirement | A goal or task that specific classes of users must be able to perform with a system, or a desired product attribute. | *Goal* | |

The classification of this article appears to cover the Wiegers-Beatty categories.

## 7.2 Van Lamsweerde

In the same style as 7.1, the following table considers the classification by [van Lamsweerde 2019] ('AVL") from which the first two columns are reproduced verbatim.

| AVL category | AVL definition | Category from the present discussion | Comments |
|---|---|---|---|
| Functional requirements | Functional effects that the software-to-be is required to have on its environment. | *Constraint* or *Behavior* | |
| Non-functional requirements | Constraints on the way the software-to-be should satisfy its functional requirements or on the way it should be developed. | *Task* | Can also be *product* |
| Quality requirements | Additional, quality-related properties that the functional effects of the software-to-be should have. | *Constraint* | Usually *engineering decisions* |
| Compliance requirements | Prescribed software effects on the environment to conform to national laws, international regulations, etc. | *Constraint* | Usually *business rule* |
| Architectural requirements | Imposed structural constraints on the software to fit its environment. | *Component* | |

| Development requirements | Non-functional requirements on the way the software-to-be should be developed. | *Task* | Can also be *product* |
|---|---|---|---|

The following artifacts are not defined as requirements categories in [van Lamsweerde 2019], but are important enough for inclusion here:

| Goals | Prescriptive statements of intent that the system should satisfy through the cooperation of its agents (active system components). | *Goal* | |
|---|---|---|---|
| Expectations | Goal under the responsibility of a single agent in the environment of the software-to-be. | *Goal* | |
| Domain properties | Descriptive statement about the environment, expected to hold invariably regardless of how the system behaves. | *Constraint* | Or *Component* if the property holds on a structural description. |

Coverage again appears good.

# 8. Normative work

This section considers some existing normative work on requirements.

## 8.1 IEEE definition

The current version of the IEEE standard for software terminology [IEEE 2010], released in 2010, offers a definition of "requirement", retained and confirmed from a 1990 version. Under that definition, a requirement is:

> *1. A condition or capability needed by a user to solve a problem or achieve an objective.*

> *2. A condition or capability that must be met or possessed by a system, system component, product, or service to satisfy an agreement, standard, specification, or other formally imposed documents.*

> *3. A documented representation of a condition or capability as in (1) or (2).*

> *4. A condition or capability that must be met or possessed by a system, product, service, result, or component to satisfy a contract, standard, specification, or other formally imposed document. Requirements include the quantified and documented needs, wants, and expectations of the sponsor, customer, and other stakeholders.*

That definition cannot be right. Its very length is just a symptom of the problem: "requirement", either in ordinary usage or as applied to software, is a simple concept which merits a simple definition.

In clause 1, a requirement is a "*condition or capability*", but it is not clear what these terms mean and how the meanings differ; "capability" is not defined in the standard, and "condition" is defined as "*a description of a contingency to be considered in the representation of a problem, or a reference to other procedures to be considered as part of the condition*", where "contingency" is not defined. This definition of "condition" is indefensible: it is again far too complex and mysterious, especially in light of the ordinary-language meaning of the term (as everyone knows, a condition is simply, a property that can be true or false). That ordinary meaning would seem just right in a systems/software context too. Coming back to the definition of "requirement", the distinction between *"solve a problem"* or *"achieve an objective"* seems spurious (solving a problem is an objective, and reaching an objective raises problems).

The distinction between clause 2 and clause 1 is equally uninteresting, since the definition of "*user*" in the standard, too long (18 lines!) to be reproduced here, is broad enough to encompass anyone having an interest in an agreement, standard etc. Worse, clause 2 makes the definition circular, since a "specification" (defined as "*a detailed formulation, in document form, which provides a definitive description of a system for the purpose of developing or validating the system*") certainly includes the description of all "conditions" and "contingencies" of the system, whatever those may be; so a requirement is defined as a condition that must be met to satisfy a specification of conditions!

Viewed in light of the distinction between a *property* and a *statement* of that property (section 3.2), clause 3 commingles these two notions under the term "requirement", a source of confusion: a property is not the same thing as one representation of that property in some notation such as English, UML or Telugu.

Clause 4 is entirely mystifying, since it is almost identical to clause 2 but not quite, raising issues of consistency; in addition, the commingling of property and statement of clause 3 does not apply to clause 4, leaving the reader wondering.

As to the last sentence, it is not in the form of a definition like the preceding ones, but comments on what requirements may "include"; such sentences, inappropriate in a definition since they can only serve to confuse the reader further (if the first four clauses, already lengthy and redundant, are supposed to define requirements, what else is needed?); it sounds more like a "remorse", a typical flaw of definitions [Meyer 1985-2], trying to make up for an unsatisfactory definition by adding a broad net of precautionary qualifications at the end.

Insistent as it is on including irrelevant and redundant details, the definition manages to miss crucial aspects of requirements: it focuses on system requirements, but does not cover properties of the *project*, and may cover *environment* properties only by a stretch of the imagination.

This addled attempt at a definition, which sounds like an attempt to integrate the comments of everyone in a committee, is unlikely ever to have helped a software practitioner. One should note here that such self-defeating pomposity is inevitable neither for standards in general nor for IEEE standards. The 1998

IEEE requirements standard [IEEE 1998], long marked as obsolete but still widely used in the industry (which prefers it to its successors, an understandable attitude in light of the present discussion's example), is a short, clear, no-frills standard, and as a result remarkably useful in practice.

The IEEE-2010 definition does have one redeeming feature: its restriction to properties "*needed by a user*". Through this clause, the definition expresses that not all properties (of a project, system or environment) are interesting as requirements only if they are of interest to someone. That someone should be defined not as "*a user*" but as a stakeholder. (Many legitimate requirements are intended for stakeholders other than users, for example to company management in the case of requirements that the present discussion classifies as *goals*. A goal such as "take market share away from competitor X" is is relevant as a requirement, but hardly "*needed by a user*". It is needed by a stakeholder. This sloppiness in terminology is all the more surprising that the standard does define "stakeholder".) Still, the underlying idea is correct: a requirement is not just any property of the system (or project, or environment) but one that some stakeholder (e.g. a user) finds important. The present article's definition of requirement recognizes that idea by defining the concept of a *relevant* property (3.3) and including it in the definition of "requirement" (3.4).

## 8.2 SWEBOK

SWEBOK, the IEEE-originated Software Engineering Book of Knowledge [Bourque 2014], is an effort to classify existing knowledge in software engineering, with numerous elements in common with the IEEE standard discussed above.

SWEBOK defines a "requirement" as "*a property that must be exhibited by something in order to solve some problem in the real world*". This definition is in part useless and in part wrong:

- It is grammatically challenged. As written, it implies that it is the "property" that must "solve some problem". Since properties do not solve problems, the most reasonable interpretation, which we will assume, is that the definition is incorrect English for "... in order *for someone* to solve some problem". This point of pure form is not just quibbling since a definition, particularly in a document attempting to define best practices, is only useful if it is clear.
- On the substance: why the "real" world? What would be a "problem" in an *unreal* world? "Real world" is informal language, not a concept for a standard of industrial practice. SWEBOK uses it more than a dozen times but does not define it. The intention seems to be that software should not exist just for itself, and instead should be related to some issue in the non-software world, like banks or airplanes. But this view, while common in simplistic discussions of software engineering, is incorrect: requirements are defined and necessary for systems that are entirely virtual and not part of the physical world, like a compiler, an operating system, a Web browser…
- While too restrictive in its focus on the "real world", the definition is too general in other ways. "In order to solve" the "problem" of building a software system, a "property" that must be "exhibited by" the building hosting the team ("something") is that it should not be on fire, and a property of the team members (another "something") is that they should be awake. Those are hardly requirements in any meaning pertaining to software engineering.

After this useless definition, SWEBOK introduces some more relevant concepts, such as "product requirement" and "process requirement" which, tellingly, are defined without reference to it: respectively, "*need or constraint on the software to be developed*" and "*essentially*" (?) "*a constraint on*

*the development of the software*". The first of these definitions seems to confuse behaviors and constraints, since it is illustrated by the example "*The software shall verify that a student meets all prerequisites before he or she registers for a course*". Such a property is not "a need or constraint **on** the software" (which would be something like "registration to a course is conditional on satisfying the prerequisites", an environment property) but a property *of* the software (a behavior in the terminology of the present work). The fundamental distinction between properties of the environment and properties of the system is one of the insights gained in the progress of software engineering over the past two decades, but SWEBOK is not aware of it, other than in a brief mention of "business rules" in the section on requirements elicitation.

As these samples illustrate, SWEBOKS's strength is not in definitions of software engineering concepts, or more generally in precision and clarity (all the more regrettable that many textbooks reverently cite SWEBOK as a font of software engineering wisdom). It naturally tends to the prescriptive mode and includes (aside from such time-wasting platitudes as requirements elicitation being "*fundamentally a human activity*") some reasonable advice, such as ensuring "*effective communication between the stakeholders*" to guarantee good requirements elicitation.

The aspect of SWEBOK most relevant to the present effort at taxonomy is the attempt at requirements classification along "*a number of dimensions*": functional vs nonfunctional, single versus emergent, product versus process, higher or lower priority, scope, volatility versus stability.

## 8.3 Essence

Essence [OMG 2018], by the Semat consortium under the leadership of Ivar Jacobson, is an effort to develop a systematic understanding of software engineering concepts and best practices. Requirements appear as one of seven "alphas" (key elements) of Essence, along with Software System, Team, Work, Way of Working, Opportunity ("*The set of circumstances that makes it appropriate to develop or change a software system*") and Stakeholders. Essence defines the role of requirements as "*what the software system must do to address the opportunity and satisfy the stakeholders*". This definition is indefensible since it covers only one of the three relevant aspects, the system (3.4), missing the project and the environment. (It fails to cover such typical requirement examples "version 1 shall be operational no later than September 2023" and "the social security number uniquely identifies a person", respectively project and environment properties.)

Like many software engineering discussions, Essence does not devote much effort to defining basic concepts and instead veers quickly into prescriptive mode. In fact, immediately after the preceding definition comes the prescriptive observation that "*It is important to discover what is needed from the software system, share this understanding among the stakeholders and the team members, and use it to drive the development and testing of the new system.*" The main contribution of the Essence discussion of requirements is indeed prescriptive: defining a sequence of states through which requirements progressively become more mature, including successively:

- Four states relative to the requirements just by themselves: Conceived (need for a new system agreed), Bounded (purpose is clear), Coherent (consistent description of system essentials), Acceptable (requirements are satisfactory for stakeholders).
- Two states that also involve the implementation: Addressed (enough to satisfy the need for a new system); Fulfilled (fully satisfies stakeholders).

Could Essence contribute to the present effort at taxonomy? Unfortunately (and surprisingly for such a recent effort) Essence suffers from the same dated view of requirements as SWEBOK, not integrating the progress of its understanding over the last two decades. The basic definition, as noted above, covers only the system part. Interestingly, the notion of environment does appear, but only twice and without explanation, in the description of the Bounded state ("*constraints are identified and considered*" and "*assumptions are clearly stated*"). There is no mention of project aspects, other than a condition in the Conceived state that "*the stakeholders that will fund the initial work on the new system are identified*". The early section on "*Justification: Why requirements?*" starts: "*the requirements capture what the stakeholders want from the system*"; this view is naïve since the requirements for a practical system requirements cannot just consider what the stakeholders want but also what is possible. In fact, out of the nine basic categories of requirements from 4.1 (ignoring meta-requirements), an SRS capturing only "what the stakeholders want" would only cover one, goals, and possibly part of another, behaviors.

Essence does introduce a concept useful to the discussion of requirements: one of the alphas, "opportunity" defined (as noted) as "*the set of circumstances that makes it appropriate to develop or change a software system*". In relation to the present work's terminology, an opportunity is the basic reason behind a goal. For example, if one of the goals of a project (back in the late 1990s) was "make our billing system ready for the transition to the Euro", that goal only made sense because of the opportunity, in the Essence meaning, that some European countries are replacing their separate currencies by a common one. For the discussion of requirements, this notion is one level too far from software development: a software system does not directly "*address the opportunity*", as the Essence definition of requirements (cited above) says: it addresses a goal. Between the switch to the euro, an opportunity in Essence terms, and the software update, a *system* effort, stands a *goal*: adapt the software to be ready for the switch. The goal addresses the opportunity; the requirements address the goal. Still, by highlighting the concept of opportunity Essence reminds us that in the broader context of software engineering behind every goal stands an opportunity.

The six stages in the Essence progression of requirements are also an interesting contribution, but they belong to the prescriptive realm beyond the scope of the present work.

The other way around in the relationship, we suggest that future versions of Essence could take advantage of the present work. Essence is a commendable effort to establish software engineering on a more solid basis, but cannot reach this goal without precise definitions (which, as we saw, industry standards do not provide) of the core concepts. In the case of requirements it needs to be brought in line with the modern understanding of these concepts.

## 9. Assessment and future work

The expected contributions of this work include providing a basis for:

1. Clarifying requirements concepts, through precise, non-bureaucratic, non-pompous but effectively usable definitions.
2. Requirements methodology ("prescriptive" discussions of requirements).
3. The critical analysis of requirements documents, as part of a quality assurance and improvement process.
4. Automatic processing of natural-language requirements documents.
5. Formal approaches to requirements (as discussed in a survey [Bruel 2019]).

On point 2, we may note that much of the existing literature on requirements is prescriptive: textbooks tell students what distinguishes good requirements from bad, and research articles propose new requirements methods meant to improve on existing practices. This focus is understandable, particularly since it is a widely shared assessment that the quality of requirements as actually written in industry is overall not very good. The present work is at a different, more basic level: providing fundamental definitions and taxonomies to enable better understanding and discussions of requirements. As one of its applications, it can help inform prescriptive discussions, and make them more effective, by defining the framework precisely. We saw some examples of possible prescriptive consequences of the descriptive approach of this work:

- The distinction between *homogeneous* and *heterogeneous* composite requirements (section 3.2) leads to the observation that the second kind is to be avoided. If a requirement is composite, it should bind together sub-requirements of a similar nature and not, for example, a component and a behavior, or a behavior (applying to the system) and a constraint (characterizing the environment).
- The notion of *component* is closely connected to the advice (present in all good requirements methods, going back to the venerable IEEE standard on requirements [IEE 1998]) to list and define all relevant concepts in a glossary. All important components should appear in the glossary.
- The notion of *lack* directs requirements engineers and quality assurance teams to look for requirement elements that have been overlooked. An example of lack is a component that does not appear in the glossary.
- The notion of *contradiction* again provides guidelines for quality assurance on requirements. Practical requirements document often contain a surprising number of contradictions, arising in particular from long periods of requirements development and the intervention of many different people in the process.
- The notion of *repetition* (REPEATS relation) is also important, in particular when distinguishing between two of the relation's variants: EXPLAINS is legitimate (provide different views of the same property, in different notations), although it is important to ensure consistency as in the "multirequirements" approach [Meyer 2013]; DUPLICATES, on the other hand, is in our view always bad. (One could state that repeating the same information in different ways but in the same notation can be harmless, but it is not: the duplication contributes to requirements document bloat; it wastes the reader's time; it can confuse the reader who does not know which of different explanations of the same property to believe; and it fares poorly in the context of software evolution since it is easy to update one variant and forget the others.)
- The important recurring debate between traditional ("waterfall") and agile approaches to requirements can benefit from the precise analyses of the present work.

On points 3 and 4 (analysis of SRS), the precision that we have tried to apply to the definitions and taxonomies should help efforts to perform automatic NLP (Natural-Language-Processing) analysis of requirements document. There has been considerable research interest in this topic. NLP and more generally AI techniques have made astounding advances, but they are better at inferring a good-enough approximation of a considerable amount of information than at inferring precise information. An example (hijacked from a discussion of agile methods in [Cohn 2010]) is, in a requirements specification for a seminar scheduling system, the property that "the hotel is booked": it could mean that we have just succeeded in booking the hotel, or that it was already booked by someone else and hence that we have to look for another. While humans can handle this kind of subtlety, it seems beyond the reach of algorithms. But automatic analysis does not raise that level of difficulty if it focuses on structure rather than deep semantics. Its goal then is to organize the requirements, decode ("parse") the structure of the project, system and environment, and identify relations. Such an analysis could yield a first level of formalization of informal requirements, useful by itself (and also as a starting point for finer semantic analysis, automatic or partly manual). Building the corresponding tools, by relying on the concepts developed in this article, seems a promising avenue of research with achievable goals.

Such NLP processing based on the taxonomies of this article is part of our current work. Other efforts in progress include:

- Exploring properties of requirements in relation to other software artifacts, such as code, whereas the present discussion mostly considers requirements by themselves.
- Validating the approach on many further examples, academic and industrial.
- Assessing its teachability, by using it in courses on software engineering and requirements.
- Using it as a basis for a formal specification of requirements concepts. There have been various attempts to describe software engineering concepts in formal frameworks. (An early example was ([Meyer 1985], which provides a mathematical model for binary relations between program elements such a modules, expressing formal properties of these relations.) The present discussion provides a solid basis for discussing requirements concepts, but it is still expressed in natural language rather than mathematics. We believe it provides an excellent starting point for mathematical modeling of the concepts under discussion and hope to develop the corresponding formal specifications, with a view to uncovering laws of software engineering that admit rigorous mathematical statements.

Even without these further developments, we hope to have provided a clearly defined framework that can serve as a reference for future work on requirements, and help improve the state of the art in this critical area of software engineering.

## Acknowledgments

We are grateful to Dr. Bettina Bair from Ohio State University for writing the original (2006) version of the course project document [Bair 2006] and providing us with a more recent version.

Attendees of talks given on this work by some of the authors provided particularly relevant feedback: at Politecnico di Milano (Meyer, March 2019), Elisabetta Di Nitto, Carlo Ghezzi, Dino Mandrioli and Maurizio Patriarca; at the University of Toulouse (Meyer, March 2019), Mamoun Filali Amine, whose comments led to a revision of the classification of constraints; at Innopolis University (Meyer, March 2019); at the GDR meeting, Génie de la Programmation et du Logiciel, also in Toulouse (Bruel, June 2019).

We are further indebted to Joëlle Guion for important comments on the concerns of practicing requirements engineers.

### Revision history
2 December 2018 Paper draft (Meyer as sole author). Other authors join.

March-June 2019: talks by Meyer and Bruel (Politecnico di Milano, Univ. of Toulouse, Innopolis University, GDPR conference); integration of attendees' comments. Design and implementation of online quizzes by Galinier. Naumchev joins as co-author.

16 June 2019: first arXiv version submitted (published 18 June).

28 June 2019: second arXiv version submitted (this version). Correction of mistakes, addition of SWEBOK (8.1) and Essence sections (8.2). Addition of FOLLOWS relation.